\documentclass[journal=jpclcd]{achemso}

\usepackage[T1]{fontenc} 

\usepackage{amsmath,color}
\usepackage{amssymb}
\usepackage{amsfonts}
\usepackage{amsthm}
\usepackage{mathtools}

\usepackage{algorithm}
\usepackage{algpseudocode}
\usepackage{dsfont}

\usepackage{bbold}
\usepackage{chemformula}
\usepackage{siunitx}

\usepackage{pdfpages}

\DeclarePairedDelimiterX\braket[2]{\langle}{\rangle}{#1 \delimsize\vert #2}
\DeclarePairedDelimiterX\braket3[3]{\langle}{\rangle}{#1 \delimsize\vert #2 \delimsize\vert #3}

\usepackage{accents}

\usepackage{fancyvrb}

\newcommand{\vk}{\mathbf{k}}

\newcommand{\vxi}{\boldsymbol{\xi}}

\newcommand{\hH}{\hat{H}}

\newcommand{\avg}[1]{\left\langle #1\right\rangle}

	\title{Quantum Simulations of Vibrational Strong Coupling via Path Integrals}
	
	\author{Tao E. Li}%
	\email{tao.li@yale.edu; taoli@sas.upenn.edu}
	\affiliation{Department of Chemistry, Yale University, New Haven, Connecticut, 06520, USA}
	\affiliation{Department of Chemistry, University of Pennsylvania, Philadelphia, Pennsylvania 19104, USA}
	
	\author{Abraham Nitzan} 
	\email{anitzan@sas.upenn.edu}
	\affiliation{Department of Chemistry, University of Pennsylvania, Philadelphia, Pennsylvania 19104, USA}
	\affiliation{School of Chemistry, Tel Aviv University, Tel Aviv 69978, Israel}
	
	\author{Sharon Hammes-Schiffer}%
	\email{sharon.hammes-schiffer@yale.edu}
	\affiliation{Department of Chemistry, Yale University, New Haven, Connecticut, 06520, USA}
	
	\author{Joseph E. Subotnik}
	\email{subotnik@sas.upenn.edu}
	\affiliation{Department of Chemistry, University of Pennsylvania, Philadelphia, Pennsylvania 19104, USA}

\begin{document}

    \begin{tocentry}
		\includegraphics[width=1.\linewidth]{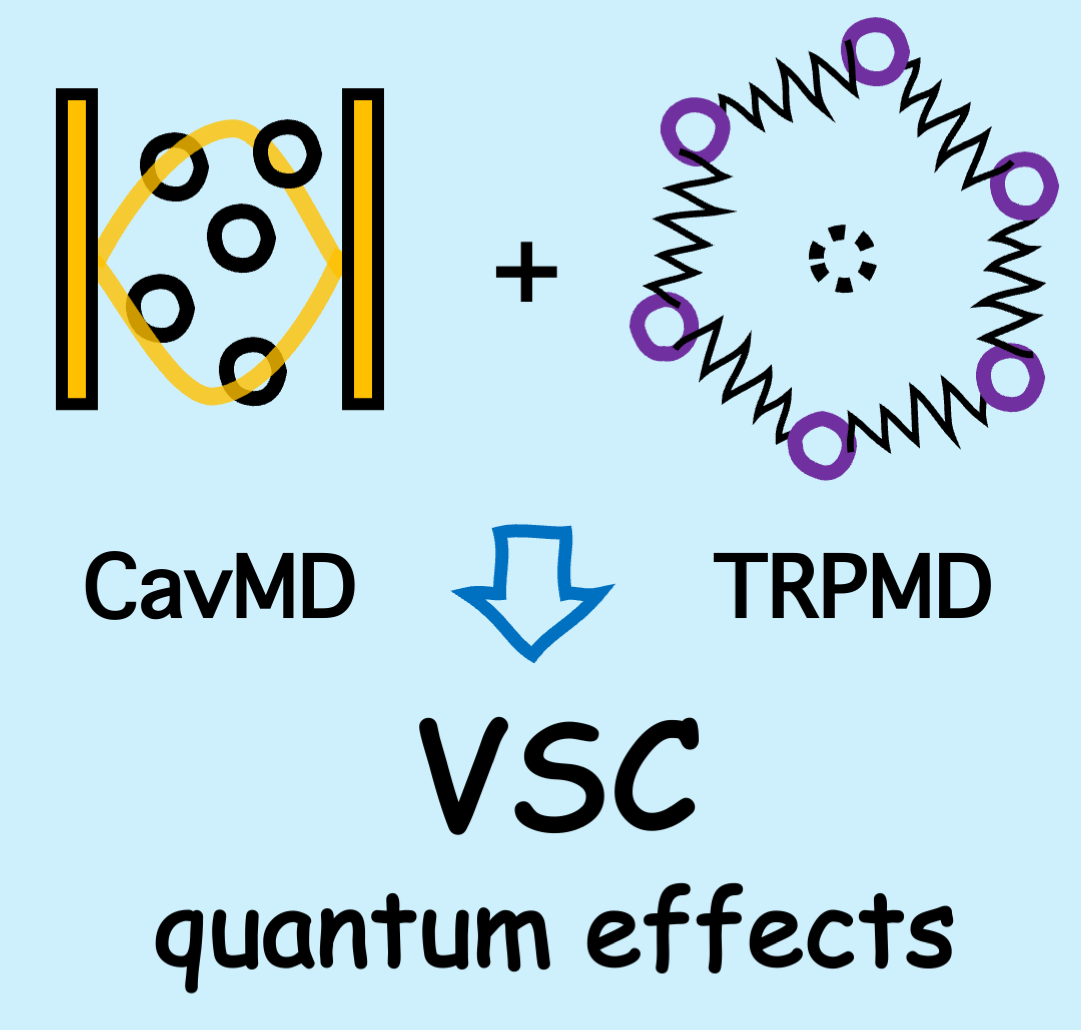}
	\end{tocentry}
	
	\begin{abstract}
	    A quantum simulation of vibrational strong coupling (VSC) in the collective regime via thermostatted ring-polymer molecular dynamics (TRPMD) is reported. For a collection of liquid-phase water molecules resonantly coupled to a single lossless cavity mode, the simulation shows that, as compared with a fully classical calculation, the inclusion of nuclear and photonic quantum effects does not lead to a change in the Rabi splitting but does broaden polaritonic linewidths roughly by a factor of two. Moreover, under thermal equilibrium, both quantum and classical simulations  predict that the static dielectric constant of liquid water is largely unchanged inside versus outside the cavity. This result disagrees with a recent experiment demonstrating that the static dielectric constant of liquid water can be resonantly enhanced under VSC,  suggesting either limitations of our approach or perhaps other experimental factors that have not yet been explored.
	\end{abstract}

	\maketitle
	
	
	In the presence of strong light-matter interactions, molecular properties can be modified by forming hybrid light-matter states, known as  polaritons. \cite{Ribeiro2018,Flick2018,Herrera2019,Forn-Diaz2019,FriskKockum2019,Xiang2021JCP,Garcia-Vidal2021,Li2022Review} Recently, a great deal of  interest has been focused on the vibrational strong coupling (VSC) regime, \cite{Long2015,George2015} in which the bright collective mode of molecular vibrations forms a pair of upper and lower polaritons (UP and LP) with an optical cavity mode in a Fabry--P\'erot microcavity. While experiments have shown that forming VSC can resonantly modify ground-state chemical reaction rates, \cite{Thomas2016,Thomas2019_science,Lather2019,Pang2020,Sau2021,Lather2021,Vergauwe2019} crystallization processes, \cite{Hirai2020Crys} and supramolecular assembly, \cite{Joseph2021} a theoretical understanding of VSC \cite{Galego2019,Campos-Gonzalez-Angulo2019,Li2020Origin,Campos-Gonzalez-Angulo2020,LiHuo2021,Sidler2021,DuNoneq2021,Schafer2021} is still lacking.
	
	One route toward understanding VSC is to perform numerical simulations for realistic molecules inside the cavity.\cite{Triana2020Shape,Li2020Water,LiHuo2021,Schafer2021} Along this direction, we have developed a classical cavity molecular dynamics (CavMD) scheme. \cite{Li2020Water} 
	This approach self-consistently propagates the coupled classical dynamics between a few cavity modes and a large ensemble of  condensed-phase molecules moving on an electronic ground-state surface.
	Equipped with this approach, we have carefully scrutinized some fundamental VSC processes and gained understanding about existing \cite{Li2020Water,Li2021Relaxation,Li2020Nonlinear,Li2021Collective} and potentially interesting \cite{Li2021Solute} experiments. However, one outstanding question remains: to what extent can we trust {\em classical} simulations? Because the molecular vibrational frequencies and cavity frequencies involved in VSC are usually much larger than room temperature (300 K $\sim$ 209 cm$^{-1}$), quantum effects cannot be simply ignored.

	Here, in order to investigate the potential impact of nuclear and photonic quantum effects, we report the first quantum CavMD simulation of liquid water under VSC  by path-integral techniques. \cite{Parrinello1984,Craig2004,Habershon2013,Markland2018}   While path-integral techniques have been used to study strong light-matter interactions, \cite{Li2020Origin,Chowdhury2021,YangCao2021}  a realistic simulation in the collective regime 
	has not yet been reported. Due to the importance of liquid water, understanding how nuclear and photonic quantum effects alter the VSC dynamics  \cite{Vergauwe2019,Hiura2018,Imperatore2021,Fukushima2021water,Fukushima2021Prom} is fundamentally intriguing. Apart from this fundamental motivation, performing a quantum simulation is also needed for a better understanding of the gap between recent VSC experiments  and theory. For example, a recent VSC experiment in liquid water \cite{Fukushima2021Prom} shows that under thermal equilibrium, proton conductivity and the static dielectric constant of liquid water can be resonantly enhanced under VSC.  At the same time, however, our previous analytical studies \cite{Li2020Origin} and classical CavMD simulations \cite{Li2020Water} have demonstrated that all equilibrium properties of liquid water should be practically the same in and out of the cavity. Given that the static dielectric constant is also an equilibrium property, \cite{Paesani2006} there is clearly an inconsistency between our previous classical simulations and experiments. Hence, a quantum simulation becomes necessary to better understand this conflict. Moreover, because a hypothetical resonant effect for the dielectric constant would be similar to resonant effects in VSC catalytic experiments, \cite{Thomas2016,Thomas2019_science,Lather2019,Pang2020,Sau2021,Lather2021,Vergauwe2019} understanding whether or how VSC induces a resonant enhancement of the static dielectric constant may be very important for interpreting VSC catalytic experiments.
	
	
	Before demonstrating the results, we briefly outline the fundamentals of path-integral techniques and CavMD.
	Path-integral molecular dynamics (PIMD) can accurately calculate quantum equilibrium properties of distinguishable particles at finite temperatures, \cite{Tuckerman2010} for which nuclei of molecules moving in an electronic ground state are a good approximation. Ring-polymer molecular dynamics (RPMD) \cite{Craig2004,Habershon2013,Markland2018} extends PIMD and can also approximate quantum dynamical properties of molecules reasonably well, especially for  condensed-phase systems at finite temperatures, where dephasing processes can easily destroy quantum entanglement and make the molecular system more "classical". In PIMD and RPMD, quantum dynamics of molecules are propagated in an extended classical phase space, in which $P$ classical molecular trajectories, also known as $P$ beads, interact with each other through a harmonic interbead interaction. Due to this interbead  interaction, nuclear quantum effects, including zero-point energy, \cite{Habershon2009ZPE} delocalization, and quantum tunneling, can be efficiently described.
	However, due to spurious resonances among the normal modes of the beads, RPMD can sometimes fail to accurately describe the molecular spectrum in the high-frequency domain. \cite{Habershon2008}
    One simple means to solve this issue is to perform thermostatted RPMD (TRPMD),\cite{Rossi2014,Ceriotti2010} in which higher frequency internal modes of the ring polymer are attached to an additional thermostat. For our simulations, because the Rabi splitting in the high-frequency domain is critical for VSC, we will perform TRPMD instead of the standard RPMD. A more detailed description of RPMD is also given in the SI Sec. I.
	
	After introducing TRPMD, let us  briefly review the fundamentals of CavMD. Within the framework of CavMD, the quantum electrodynamical (QED) Hamiltonian on the electronic ground state is \cite{Li2020Origin,Li2020Water,Flick2017}
	\begin{subequations}\label{eq:H_QED}
	\begin{equation}\label{eq:H_QED-1}
	    \hH_{\text{QED}}^{\text{G}} = \hH_{\text{M}}^{\text{G}} + \hH_{\text{F}}^{\text{G}},
	\end{equation}
	where  $\hH_{\text{M}}^{\text{G}}$ denotes the standard (kinetic + potential) molecular Hamiltonian, and the field-related Hamiltonian $\hH_{\text{F}}^{\text{G}}$ is 
	\begin{equation}\label{eq:H_QED-2}
	    \hH_{\text{F}}^{\text{G}} = \sum_{k,\lambda} \frac{\hat{\widetilde{p}}_{k,\lambda}^2}{2m_{k,\lambda}}
	    + \frac{1}{2}m_{k,\lambda}\omega_{k,\lambda}^2
	    \left(
	    \hat{\widetilde{q}}_{k,\lambda} + \frac{\hat{d}_{\text{g},\lambda}}{\omega_{k,\lambda}\sqrt{\Omega \epsilon_0 m_{k,\lambda}}}
	    \right)^2.
	\end{equation}
	\end{subequations}
	Here, $\hat{\widetilde{p}}_{k,\lambda}$, $\hat{\widetilde{q}}_{k,\lambda}$,  $\omega_{k,\lambda}$, and $m_{k,\lambda}$ denote the momentum operator, position operator, frequency, and auxiliary mass for each cavity photon mode with wave vector $\vk$ ($k = |\vk|$) and polarization direction defined by a unit vector $\vxi_\lambda$ (with $\vxi_\lambda \cdot \vk = 0$). $\hat{d}_{\text{g},\lambda}$ denotes the ground-state molecular dipole operator of the whole molecular subsystem projected along the polarization direction $\vxi_\lambda$. In many cases,  $\hat{d}_{\text{g},\lambda} = \sum_{n=1}^{N} \hat{d}_{n\text{g},\lambda}$ can be expressed as a summation of the dipole operators of each individual molecule, where $N$ denotes the total molecular number. $\Omega$ denotes the cavity volume and $\epsilon_0$ denotes the vacuum permittivity. Although Eq. \eqref{eq:H_QED-2} has a summation over many cavity modes, in practice, during simulations, we will take into account only one cavity mode (with two polarization directions) that is at resonance with the molecular vibrational mode.

	Clearly, in the QED Hamiltonian defined by Eq. \eqref{eq:H_QED}, the cavity photons can be regarded as additional "nuclear" degrees of freedom of the molecular system, and the interaction potential between cavity photons and molecules (the last term in Eq. \eqref{eq:H_QED-2}) depends on the position operators only. In other words, Eq. \eqref{eq:H_QED} resembles a standard (kinetic + potential) molecular Hamiltonian, so we can directly apply TRPMD to calculate quantum equilibrium and dynamical properties of both the molecules and the cavity photons in the same manner as TRPMD for the molecules outside the cavity. In our implementation, the nuclei and cavity photons are each represented by 32 beads ($P=32$). 
	In order to distinguish between these simulations and TRPMD simulations outside the cavity, we will often refer to our generalized TRPMD simulations under VSC as  quantum CavMD simulations. 
	Note that when the number of beads becomes unity ($P=1$), the  quantum  CavMD approach is reduced to classical CavMD.

	
	For our simulation of liquid water under VSC,  the molecular system is described with the empirical force field q-TIP4P/F. \cite{Habershon2009} This force field is commonly used in the path-integral community and can roughly generate experimentally comparable nonreactive properties of liquid water, including IR spectroscopy, the diffusion constant, and the static dielectric constant. \cite{Liu2016,Habershon2009} The liquid water system is represented by 216 \ch{H2O} molecules in a cubic simulation cell under periodic boundary conditions at 300 K. Inside the cavity, this molecular system is coupled to a lossless cavity mode (with two possible polarization directions $x$ and $y$), and the effective coupling strength between each molecule and the cavity mode is set as $\widetilde{\varepsilon} = 4\times 10^{-4}$ a.u. See SI Sec. II for a more detailed explanation of the definition of $\widetilde{\varepsilon}$ and why our simulation can be interpreted as corresponding to Fabry--P\'erot experiments. Other simulation details are explained in the SI Sec. III. As far as the technical details are concerned, CavMD is implemented by modifying the I-PI package, \cite{Kapil2019} the nuclear forces are evaluated by LAMMPS, \cite{Plimpton1995} and the raw simulation data are available at Github. \cite{TELi2020Github}

	\begin{figure}
		\centering
		\includegraphics[width=0.5\linewidth]{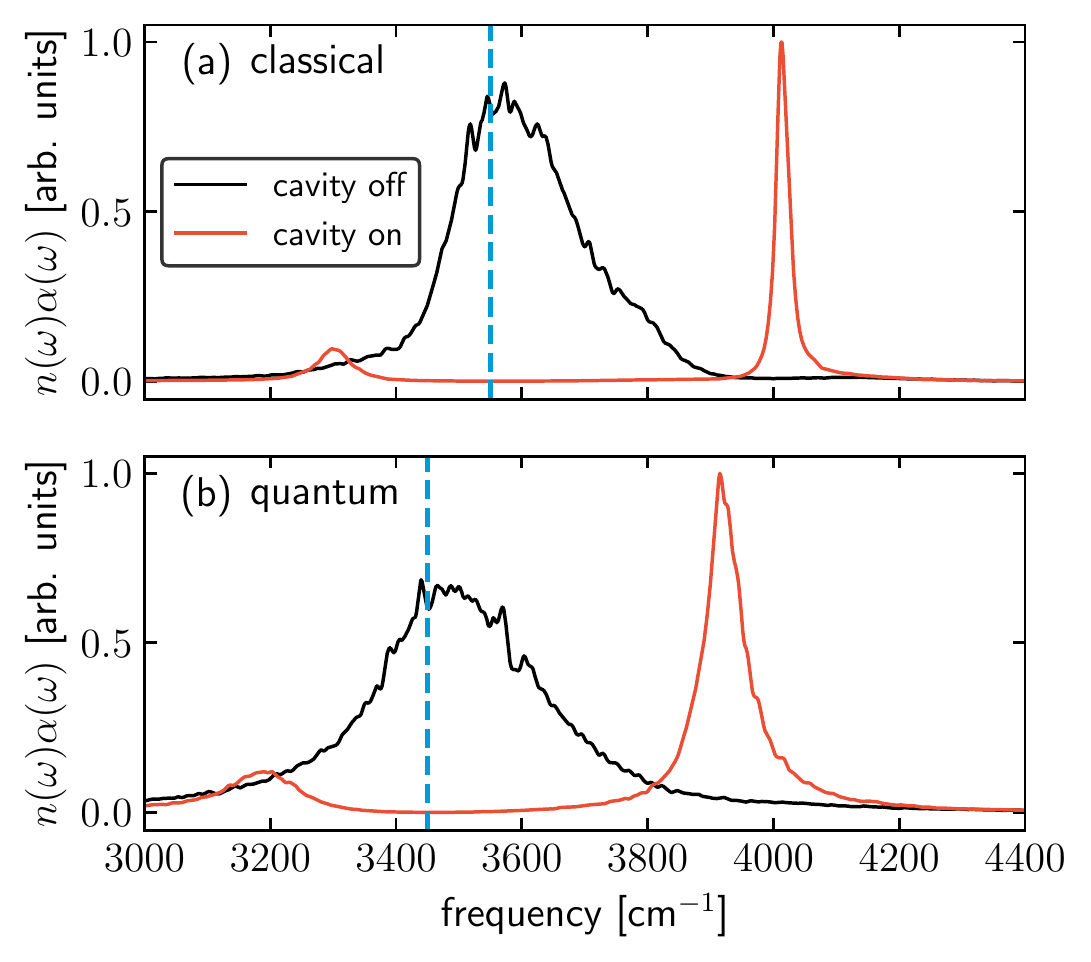}
		\caption{IR absorption spectra of liquid water from (a) classical and (b) quantum simulations. In each panel, a pair of polaritons (red line) form when the \ch{O-H} stretch band (black line) is nearly resonantly coupled to the cavity mode with $\omega_c$ = 3550 cm$^{-1}$ (classical) or 3450 cm$^{-1}$ (quantum), as indicated by the vertical dashed blue line, with $\widetilde{\varepsilon} = 4\times 10^{-4}$ a.u. Comparing the quantum results to the classical results, the Rabi splitting remains unchanged, but the polariton linewidth is significantly broadened by roughly a factor of two; see Table \ref{table:property} for the exact values.}
		\label{fig:IR}
	\end{figure}
	
	We first report the linear IR spectrum of liquid water by Fourier transforming  the dipole autocorrelation function from equilibrium  CavMD simulations. \cite{Li2020Water,Habershon2009} Outside the cavity, as shown in Fig. \ref{fig:IR}a, the classical IR spectrum (black line) shows a broad band peaking near $\omega_0 = 3550$ cm$^{-1}$, which corresponds to the \ch{O-H} stretch modes of liquid water. The full width at half maximum (FWHM) of this \ch{O-H} stretch band is $\gamma_{\text{OH}} = 215$ cm$^{-1}$. Such a large linewidth arises from inhomogeneous broadening due mainly to the static disorder of liquid water. 
	However, the classical \ch{O-H} linewidth is still considerably smaller than the experimental value of nearly 400 cm$^{-1}$. \cite{Bertie1996} Inside the cavity, when the \ch{O-H} stretch band is resonantly coupled to a lossless cavity mode with frequency $\omega_c = 3550$ cm$^{-1}$ (the vertical dashed blue line) and an effective coupling strength $\widetilde{\varepsilon} = 4\times 10^{-4}$ a.u., a pair of asymmetric UP and LP forms with a Rabi splitting of $\Omega_{\text{R}}=715$ cm$^{-1}$. Although the usual definition of ultrastrong coupling is $\Omega_{\text{R}}/2\omega_0 > 0.1$, \cite{FriskKockum2019} our value of Rabi splitting sits at the boundary between strong and ultrastrong coupling, so we will refer to our case as strong coupling only. The linewidths of the UP and LP are $\gamma_{\text{UP}} = 24$ cm$^{-1}$ and $\gamma_{\text{LP}} = 60$ cm$^{-1}$, respectively. Clearly, the summed linewidth of the LP plus the UP (84 cm$^{-1}$) is much smaller than the linewidth of the \ch{O-H} stretch band outside the cavity. This difference arises because inhomogeneous broadening of the molecular peak does not contribute to the polariton linewidth, a phenomenon that was theoretically predicted decades ago \cite{Houdre1996} and has also been experimentally confirmed under VSC. \cite{Long2015}
	
	\begin{table}
    \caption{Simulated Quantities under VSC\textsuperscript{\emph{a}}}
    \label{table:property}
    \begin{tabular}{lll}
    \hline
    VSC quantity   & Classical & Quantum  \\
    \hline
    $\omega_c \ (\approx \omega_0)$ & 3550 cm$^{-1}$  & 3400 cm$^{-1}$ \\
    $\Omega_{\text{R}}$ & 715 cm$^{-1}$  & 720 cm$^{-1}$  \\
    $\gamma_{\text{UP}}$ & 24 cm$^{-1}$ & 65 cm$^{-1}$  \\
    $\gamma_{\text{LP}}$ & 60 cm$^{-1}$  & 129 cm$^{-1}$  \\
    $\epsilon_r$ & 54.5 (54.4) & 56.6 (59.0) \\
    \hline
    \end{tabular}
    
    \textsuperscript{\emph{a}} These data were obtained from Figs. \ref{fig:IR} and \ref{fig:dielectic}, including the Rabi splitting ($\Omega_{\text{R}}$),  the LP or UP linewidth ($\gamma_{\text{LP}}$ or $\gamma_{\text{UP}}$), and the static dielectric constant ($\epsilon_r$) of liquid water. The $\epsilon_r$ values in parentheses are the corresponding outside-cavity results.  For reference, the simulated \ch{O-H} band linewidth is (classical) 215 cm$^{-1}$ or (quantum) 283 cm$^{-1}$.
    \end{table}

	Beyond the classical result, Fig. \ref{fig:IR}b plots the quantum IR spectrum of liquid water outside (black line) or inside (red line) the cavity by Fourier transforming the Kubo-transformed quantum dipole autocorrelation function. \cite{Habershon2009,Habershon2013} Outside the cavity, the \ch{O-H} stretch band peaks near 3450 cm$^{-1}$ and has a linewidth of $\gamma_{\text{OH}} = 283$ cm$^{-1}$. Compared to the classical result, this quantum peak is red-shifted by about 100 cm$^{-1}$ and is also broadened, improving agreement with the experimental values \cite{Bertie1996} compared to the classical simulations.  The red-shifting and broadening can be related to the inclusion of zero-point energy effects in TRPMD.
	Inside the cavity, when the cavity mode with frequency $\omega_c = 3450$ cm$^{-1}$ is resonantly coupled to the \ch{O-H} stretch band with $\widetilde{\varepsilon} = 4\times 10^{-4}$ a.u., the Rabi splitting between the UP and LP (red line) is $\Omega_{\text{R}} = 720$ cm$^{-1}$, in agreement with the classical result. The polariton linewidths, however, become $\gamma_{\text{UP}} = 65$ cm$^{-1}$ and $\gamma_{\text{LP}} = 129$ cm$^{-1}$, respectively. These values are significantly broadened compared with the classical results by roughly a factor of two, whereas the linewidth of  the quantum \ch{O-H} band exceeds the classical result by only 30\%. 
	
	Table \ref{table:property} further summarizes the quantum and classical values of the Rabi splitting and polariton linewidths. Clearly, the unchanged Rabi splitting shows that this quantity can be fully captured by classical mechanics. 
	As far as the lineshape is concerned, the polariton lineshape seems to be more sensitive to the quantum treatment than the molecular lineshape outside the cavity. Although one must always be aware of the limitations of TRPMD when simulating spectral lineshapes \cite{Rossi2014,Benson2019}, one can argue that the difference in the quantum and classical linewidths for polaritons reflects a quantum effect (e.g., perhaps a faster quantum polariton relaxation rate than a classical one).  
	Because polaritons are mostly harmonic under collective VSC, \cite{Campos-Gonzalez-Angulo2022} \footnote{One can understand the harmonic nature of polaritons by the following rationalization. When a polariton --- a hybrid light-matter state --- is in the second excited state, in the basis of individual molecules, the two polariton quanta can be represented by either (i)  the second excited state of each individual anharmonic molecule or (ii) two different singly excited molecules. When the number of molecules is large,  the latter scenario dominates the representation. Hence, the anharmonic nature of each individual molecular vibrational spectrum will not cause anharmonicity to the polaritonic spectrum in the collective regime.} TRPMD, a method that is expected to perform well in the harmonic limit, should describe polaritons more accurately than the anharmonic \ch{O-H} stretch band. Future work is needed to investigate the origin of the polariton broadening.

	\begin{figure}
		\centering
		\includegraphics[width=0.5\linewidth]{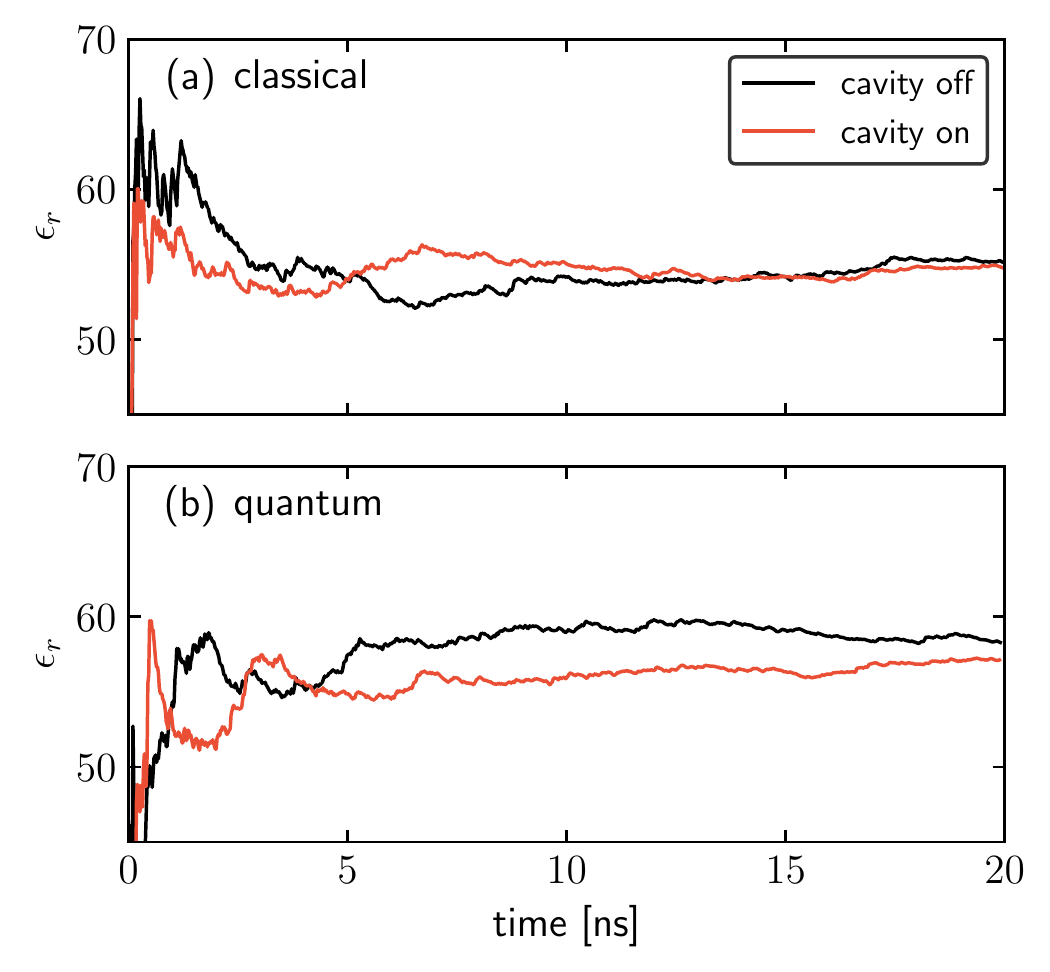}
		\caption{Static dielectric constant ($\varepsilon_r$) of liquid water as a function of time from (a) classical and (b) quantum simulations. The simulation parameters inside (red line) or outside (black line) the cavity are the same as Fig. \ref{fig:IR} except for a longer simulation time. Note that both quantum and classical simulations predict unchanged $\epsilon_r$ inside versus outside the cavity.}
		\label{fig:dielectic}
	\end{figure}
	
	Next we examine the static dielectric constant of liquid water under VSC. In general, by performing molecular dynamics simulations, the static dielectric constant ($\epsilon_r$) of a molecular system can be obtained using the following standard formula: \cite{Paesani2006}
	\begin{equation}\label{eq:dielectric_constant}
	    \epsilon_r =  1 + \frac{4\pi \beta}{3V} \left(
	    \avg{\mathbf{d}^2} - \avg{\mathbf{d}}^2
	    \right),
	\end{equation}
	where  $V$ denotes the simulation cell volume, and $\mathbf{d} = (d_{\text{g},x}, d_{\text{g},y}, d_{\text{g},z})$ denotes the  dipole moment vector of the molecular system in three dimensions.
	As a static equilibrium property, $\epsilon_r$ can be exactly evaluated by both PIMD and RPMD. However, in order to be consistent with the above IR results, we will perform TPRMD to obtain $\epsilon_r$.

	Fig. \ref{fig:dielectic}a plots the calculated $\epsilon_r$ in the time domain for up to $t = 20$ ns both outside the cavity (black line) and under VSC (red line) for the classical simulations. 
	Because the hydrogen-bonding network for liquid water restricts the relaxation of $\mathbf{d}$, in order for $\epsilon_r$ to converge, the simulation needs to run for more than 10 ns. \cite{Paesani2006,Habershon2009,Ceriotti2010} 
	Fig. \ref{fig:dielectic}b plots the analogous calculated $\epsilon_r$ inside and outside  the cavity for the quantum simulations.
	 
	In Table \ref{table:property} we report the  values of $\epsilon_r$ by taking an average between $t = 10$ ns and 20 ns. Similar to our previous observation, \cite{Li2020Origin,Li2020Water} with a classical CavMD simulation,  $\epsilon_r$, a static equilibrium property, remains unchanged inside ($\epsilon_r = 54.5$) versus outside ($\epsilon_r = 54.4$) the cavity.
	For the quantum calculations,
	the result outside the cavity ($\epsilon_r = 59.0$) agrees with a previous report. \cite{Habershon2009} Note that the values of  $\epsilon_r$ inside ($\epsilon_r = 56.6$) and outside ($\epsilon_r = 59.0$) the cavity differ slightly, but this small difference is of the same order of magnitude as the fluctuations of $\epsilon_r$ for different simulations according to the literature \cite{Habershon2009}.  Hence, we tentatively conclude that our quantum simulation does not predict a change  in $\epsilon_r$  inside versus outside the cavity.
	
	
	Our observation that the simulated static dielectric does not change in a cavity is not consistent with a recent experiment \cite{Fukushima2021Prom} claiming that $\epsilon_r$ can be resonantly enhanced by nearly 50\% under VSC and without external laser pumping.  Such a discrepancy could arise for two different reasons. First, it may indicate that our simulation does not account for crucial experimental details needed for observing the reported phenomenon. For example, there could be hidden dynamical factors or non-equilibrium fluctuations in the actual experiment that cannot be modeled with our simple setup. This possibility is related to recent experimental efforts \cite{Imperatore2021,Wiesehan2021} that failed to reproduce some VSC catalytic experiments. In other words, within the framework of a single cavity mode plus a large collection of realistic molecules under thermal equilibrium, it might simply be impossible to obtain a resonant VSC effect.
	The second reason for the discrepancy could be that our current theoretical treatment may be incomplete even within the constraints of the model above. For example, we have included only 216 \ch{H2O} molecules coupled to a single cavity mode, whereas in VSC experiments the Fabry--P\'erot cavities contain roughly $\sim 10^{10}$ molecules per mode volume and also a much more complicated cavity mode structure (i.e., the polaritonic dispersion relation) \cite{Li2022Review} than a single cavity mode. It might also be  possible that a more pronounced cavity effect could emerge numerically if more accurate \textit{ab initio} potentials were employed and/or the full cavity mode structure were simulated.   Future work is needed to investigate these possibilities. 
	
	In summary, we have reported a quantum CavMD simulation of liquid water under VSC by TRPMD. Compared with the classical results, a quantum simulation predicts no change in the Rabi splitting but predicts broadening of the polariton linewidths by roughly a factor of two. This polaritonic broadening is more significant than the broadening of the \ch{O-H} stretch band outside the cavity, and further study is needed to investigate its origin. 
	Moreover, by combining  path-integral techniques and CavMD, we have  
	demonstrated a numerical approach for distinguishing between quantum and classical VSC effects. Although the current work  does not show significant quantum effects, more intriguing VSC quantum effects might emerge when the following scenarios are studied: low-temperature dynamics, higher-order correlation functions, or nonlinear effects. These exciting directions may be investigated in the future.

	\begin{acknowledgement}
	This material is based upon work supported by the
    U.S. National Science Foundation under Grant No.
    CHE1953701 (A.N.);  US Department of Energy, Office of Science, Basic Energy Sciences, Chemical Sciences, Geosciences, and Biosciences Division under Award No. DE-SC0019397 (J.E.S.), and Air Force Office of Scientific Research under AFOSR Award No. FA9550-18-1-0134 (S.H.-S.). We thank Prof. Thomas E. Markland for insightful discussions.
    \end{acknowledgement}

	\begin{suppinfo}

    The following files are available free of charge.
    \begin{itemize}
        \item SI.pdf: Brief outline of the theory of PIMD and RPMD, brief introduction of CavMD, and the simulation details of this work.
        \item Code for reproducing this work is available at Github \url{https://github.com/TaoELi/cavity-md-ipi/}.
    \end{itemize}

    \end{suppinfo}

	
	\providecommand{\latin}[1]{#1}
	\makeatletter
	\providecommand{\doi}
	{\begingroup\let\do\@makeother\dospecials
		\catcode`\{=1 \catcode`\}=2 \doi@aux}
	\providecommand{\doi@aux}[1]{\endgroup\texttt{#1}}
	\makeatother
	\providecommand*\mcitethebibliography{\thebibliography}
	\csname @ifundefined\endcsname{endmcitethebibliography}
	{\let\endmcitethebibliography\endthebibliography}{}

	\pagenumbering{gobble}
	
	

\section*{Supplementary material}

# Supporting Information

# Quantum Simulations of Vibrational Strong Coupling via Path Integrals

Tao E. Li,*,† Abraham Nitzan,*,‡ Sharon Hammes-Schiffer,*,† and Joseph E. Subotnik*,‡

†Department of Chemistry, Yale University, New Haven, Connecticut, 06520, USA

‡Department of Chemistry, University of Pennsylvania, Philadelphia, Pennsylvania 19104, USA

¶School of Chemistry, Tel Aviv University, Tel Aviv 69978, Israel

E-mail: tao.li@yale.edu;taoli@sas.upenn.edu; anitzan@sas.upenn.edu; sharon.hammes-schiffer@yale.edu; subotnik@sas.upenn.edu

# 1. Theory of RPMD

The fundamental idea of RPMD is to propagate real-time quantum dynamics in an extended classical phase space, thus avoiding directly solving the time-dependent Schrödinger equation and allowing affordable simulation of condensed-phase systems. More technically speaking, for the electronic ground state, a quantum molecular system can be represented by a standard (kinetic + potential) quantum Hamiltonian

$$\hat{H}_{\rm M}^{\rm G} = \sum_{i}^{N_{\rm nuc}} \left[ \frac{\hat{\mathbf{P}}_i^2}{2M_i} + \hat{V}^{\rm G}(\hat{\mathbf{R}}_1, \cdots, \hat{\mathbf{R}}_{N_{\rm nuc}}) \right], \tag{S1}$$

where $\hat{\mathbf{P}}_i$, $\hat{\mathbf{R}}_i$, and $M_i$ denote the momentum operator, position operator, and mass for the $i$-th nucleus, $N_{\rm nuc}$ denotes the total number of nuclei, the superscript G denotes the electronic ground state, and $\hat{V}^{\rm G}(\hat{\mathbf{R}}_1, \cdots, \hat{\mathbf{R}}_{N_{\rm nuc}})$ denotes the multidimensional ground-state potential energy surface. According to RPMD, the quantum equilibrium and dynamical properties of molecules can be obtained by propagating the following extended classical Hamiltonian:[S1]

$$H_{{\rm M},P}^{\rm G}(\mathbf{p}, \mathbf{q}) = H_{{\rm M},P}^{\rm G,0}(\mathbf{p}, \mathbf{q}) + \sum_{k=1}^{P} V^{\rm G}(\hat{\mathbf{R}}_1^{(k)}, \cdots, \hat{\mathbf{R}}_{N_{\rm nuc}}^{(k)}), \tag{S2a}$$

where the so-called free ring-polymer Hamiltonian $H_{{\rm M},P}^{\rm G,0}$ is defined as:

$$H_{{\rm M},P}^{\rm G,0} = \sum_{k=1}^{P} \sum_{i}^{N_{\rm nuc}} \left[ \frac{[\mathbf{P}_i^{(k)}]^2}{2M_i} + \frac{1}{2} M_i \omega_P^2 \left( \mathbf{R}_i^{(k)} - \mathbf{R}_i^{(k-1)} \right)^2 \right]. \tag{S2b}$$

In Eq. (S2), $P$ copies of identical classical nuclei, known as $P$ beads, are connected with nearest neighbors by harmonic springs (the last term in Eq. (S2b)) with $\omega_P = 1/(\beta_P \hbar)$, $\beta_P = \beta/P = 1/(k_{\rm B} T P)$, and $\mathbf{R}_i^{(0)} \equiv \mathbf{R}_i^{(P)}$. This extended classical Hamiltonian contains $f = 3N_{\rm nuc} P$ Cartesian degrees of freedom. Through the interaction among the nearest-neighbor beads, nuclear quantum effects, including zero-point energy,[S2] delocalization, and quantum tunneling, can be efficiently described.

According to RPMD (and also PIMD), given a molecular operator $\hat{A}$ that depends on the molecular positions, the expectation value under thermal equilibrium can be exactly

evaluated by[S3]

$$\frac{1}{Z}\text{Tr}\left(\hat{A}e^{-\beta \hat{H}_{\text{M}}^{\text{G}}}\right) = \lim_{P\to+\infty} \langle \bar{A}_P(\mathbf{q})\rangle \tag{S3}$$

Here, $Z = \text{Tr}\left(e^{-\beta \hat{H}_{\text{M}}^{\text{G}}}\right)$ denotes the quantum canonical partition function, and

$$\bar{A}_P(\mathbf{q}) \equiv (P)^{-1}\sum_{k=1}^{P} A(\mathbf{R}_1^{(k)},\cdots,\mathbf{R}_{N_{\text{nuc}}}^{(k)})$$

denotes the centroid value of the classical variable $A$ by averaging over $P$ beads. Moreover, $\langle\cdots\rangle$ denotes the ensemble average over the extended classical phase space defined by Eq. (S2), i.e.,

$$\langle\cdots\rangle \equiv (2\pi\hbar)^{-f} Z_f^{-1} \int d^f\mathbf{p} \int d^f\mathbf{q} \cdots e^{-\beta_P H_{\text{M},P}^{\text{G}}(\mathbf{p},\mathbf{q})},$$

where the collective variables $\mathbf{p}$ and $\mathbf{q}$ denote the multidimensional momenta and positions of the $f = 3N_{\text{nuc}}P$ degrees of freedom, and $Z_f = (2\pi\hbar)^{-f}\int d^f\mathbf{p}\int d^f\mathbf{q}\, e^{-\beta_P H_{\text{M},P}^{\text{G}}(\mathbf{p},\mathbf{q})}$ denotes the canonical partition function of the classical system. Although Eq. (S3) is formally exact only when $P \to +\infty$, numerical evidence has shown that, for liquid water at room temperature, converged results can be obtained when, e.g., $P = 32$.[S4,S5]

When the dynamical properties of molecules are considered, RPMD (but not PIMD) also offers a classical-like approximation to Kubo-transformed quantum correlation functions:[S1,S6]

$$\tilde{c}_{AB}(t) \approx \langle \bar{A}_P(\mathbf{q}_0)\bar{B}_P(\mathbf{q}_t)\rangle_{\tau=0}. \tag{S4}$$

Here, $\langle\cdots\rangle_\tau$ is defined similarly as $\langle\cdots\rangle$ above:

$$\langle\cdots\rangle_\tau \equiv (2\pi\hbar)^{-f} Z_f^{-1} \int d^f\mathbf{p}_\tau \int d^f\mathbf{q}_\tau \cdots e^{-\beta_P H_{\text{M},P}^{\text{G}}(\mathbf{p}_\tau,\mathbf{q}_\tau)},$$

where $\mathbf{p}_\tau$ and $\mathbf{q}_\tau$ denote the multidimensional momenta and positions at time $\tau$. Moreover, $\bar{A}_P(\mathbf{q}_0)$ denotes the centroid value of the position-dependent variable $A$ at time 0, and $\bar{B}_P(\mathbf{q}_t)$ is defined similarly. Note that standard RPMD, i.e., an approach that simply propagates Eq. (S2) in real time, can fail to describe $\tilde{c}_{AB}(t)$ in the high-frequency regime.[S7] This

is because the higher frequency internal normal modes of the ring polymer (Eq. (S2b)) can sometimes cause nonphysical, spurious resonances with the high-frequency motion of the centroid normal mode of the ring polymer, leading to contaminated spectra in the high-frequency domain.[S7] One simple means to solve this issue is to perform thermostatted RPMD (TRPMD),[S4,S5] in which higher frequency internal modes of the ring polymer are attached to an additional thermostat. For our simulations, because the Rabi splitting in the high-frequency domain is critical for VSC, we perform TRPMD instead of the standard RPMD.

## 2. CavMD Scheme

In order to efficiently simulate the quantum QED Hamiltonian in Eq. (1) of the main text, we reduce all of the operators to classical observables. When classical dynamics (not RPMD) is considered, the equations of motion are

$$M_{nj}\ddot{\mathbf{R}}_{nj} = \mathbf{F}_{nj}^{(0)} - \sum_{k,\lambda}\left(\varepsilon_{k,\lambda}\widetilde{q}_{k,\lambda} + \frac{\varepsilon_{k,\lambda}^2}{m_{k,\lambda}\omega_{k,\lambda}^2}\sum_{l=1}^{N}d_{lg,\lambda}\right)\frac{\partial d_{ng,\lambda}}{\partial \mathbf{R}_{nj}}, \tag{S5a}$$

$$m_{k,\lambda}\ddot{\widetilde{q}}_{k,\lambda} = -m_{k,\lambda}\omega_{k,\lambda}^2\widetilde{q}_{k,\lambda} - \varepsilon_{k,\lambda}\sum_{n=1}^{N}d_{ng,\lambda}. \tag{S5b}$$

Here, $\mathbf{F}_{nj}^{(0)}$ denotes the molecular part of the force on each nuclei, i.e., the nuclear force outside the cavity; $\varepsilon = \sqrt{m_c\omega_c^2/\Omega\epsilon_0}$ denotes the light-matter coupling strength between each molecule and the cavity mode; and the subscript $nj$ denotes the $j$-th atom in molecule $n$.

Because VSC inside a Fabry–Pérot microcavity usually involves a macroscopic number (i.e., $10^9 \sim 10^{11}$) of molecules, it is very expensive if we propagate Eq. (S5) directly to simulate Fabry–Pérot experiments. Hence, we further apply periodic boundary conditions, i.e., we assume that the molecular subsystem can be divided into $N_{\text{cell}}$ identical periodic cells spanning three dimensions. In detail, for Eq. (S5) we approximate the total dipole moment of the molecules as $\sum_{n=1}^{N}d_{ng,\lambda} = N_{\text{cell}}\sum_{n=1}^{N_{\text{sub}}}d_{ng,\lambda}$, where $N_{\text{sub}} = N/N_{\text{cell}}$ denotes the number of molecules in a single cell. By further denoting $\widetilde{\widetilde{q}}_{k,\lambda} = \widetilde{q}_{k,\lambda}/\sqrt{N_{\text{cell}}}$, $\widetilde{\varepsilon}_{k,\lambda} = \sqrt{N_{\text{cell}}}\varepsilon_{k,\lambda}$, we

can rewrite the equations of motion in Eq. (S5) in a symmetric form

$$M_{nj}\ddot{\mathbf{R}}_{nj} = \mathbf{F}_{nj}^{(0)} + \mathbf{F}_{nj}^{\text{cav}}, \tag{S6a}$$

$$m_{k,\lambda}\ddot{\widetilde{\widetilde{q}}}_{k,\lambda} = -m_{k,\lambda}\omega_{k,\lambda}^2 \widetilde{\widetilde{q}}_{k,\lambda} - \widetilde{\varepsilon}_{k,\lambda}\sum_{n=1}^{N_{\text{sub}}} d_{ng,\lambda}. \tag{S6b}$$

Here,

$$\mathbf{F}_{nj}^{\text{cav}} = -\sum_{k,\lambda}(\widetilde{\varepsilon}_{k,\lambda}\widetilde{\widetilde{q}}_{k,\lambda} + \frac{\widetilde{\varepsilon}_{k,\lambda}^2}{m_{k,\lambda}\omega_{k,\lambda}^2}\sum_{l=1}^{N_{\text{sub}}} d_{lg,\lambda})\frac{\partial d_{ng,\lambda}}{\partial \mathbf{R}_{nj}} \tag{S6c}$$

denotes the cavity force on each nucleus. Because we invoke periodic boundary conditions, it is logical to redefine

$$\widetilde{\varepsilon}_{k,\lambda} = \sqrt{N_{\text{cell}}}\varepsilon_{k,\lambda},$$

to characterize the effective coupling between the cavity mode and each molecule in the periodic cell.

Because using $\widetilde{\varepsilon}_{k,\lambda}$ (instead of $\varepsilon_{k,\lambda}$) overestimates the light-matter interaction compared to that in Fabry–Pérot experiments, the CavMD simulated VSC effect on individual molecules may be artificially larger than those in Fabry–Pérot experiments. Thus, we usually need to check the dependence of the VSC effect on the molecular system size (while maintaining macroscopic observables such as the Rabi splitting and molecular density) to identify whether such a simulated effect persists in Fabry–Pérot cavities.[S8–S10] However, because the molecular IR spectra and static dielectric constant are functions of the total molecular dipole moment, i.e., the molecular bright mode, and the coupling between the molecular bright mode and the cavity mode is always the same during the system size enlarging process (due to the fixed Rabi splitting), our simulated results (with 216 molecules explicitly included) should reliably reproduce the bulk limit.

# 3. Simulation Details

For our simulations, 216 $H_2O$ molecules are placed in a cubic simulation cell with a length of 35.233 a.u., corresponding to a density of 0.997 g/cm$^3$. The $H_2O$ molecules are represented by the q-TIP4P/F force field.[S11] The long-range Coulomb interactions are taken into account by an Ewald summation. We assume that the cavity mirrors are placed along the $z$ direction, and a single cavity mode (with two polarization directions $x$ and $y$) is resonantly coupled to the O−H stretch band of liquid water. Similar to all previous CavMD work, the auxiliary mass of the cavity photon is set as $m_c = 1$ a.u.

For the classical simulations,[S12] in order to fully equilibrate the system, a 150 ps simulation is performed in a canonical (NVT) ensemble at 300 K, and all particles (nuclei + photons) are attached to a Langevin thermostat with a lifetime of 100 fs. After the NVT simulation, 40 consecutive microcanonical ensemble (NVE) trajectories of length 20 ps are simulated. At the beginning of each NVE trajectory, the initial velocities of all particles (nuclei + photons) are resampled according to the Maxwell–Boltzmann distribution at 300 K. The time step of the classical simulations is 0.5 fs, and the particle coordinates are stored every 2 fs. The IR spectrum is calculated by averaging over the 40 NVE trajectories. In order to calculate the dielectric constant, after the above NVT simulation, a 20 ns NVE simulation is further obtained, and the total dipole moment vector of the molecular system is printed out every time step.

The TRPMD simulations are largely the same as the classical simulations, and only the differences are mentioned here. Following Ref. S5, in order to obtain a converged result, we initialize the ring polymer with 32 beads ($P = 32$) for all nuclei and cavity photons and use a time step of 0.25 fs. In order to remove the spurious resonances in standard RPMD, a path integral Langevin equation (PILE) thermostat is attached to the internal modes of the ring polymer with an optimally damped coefficient $\lambda = 1/2$.[S5] For the centroid mode of the ring polymer, no thermostat is applied, which is valid for sampling equilibrium properties in liquid water.[S4] This PILE thermostat is used throughout all TPRMD simulations, including both the equilibration procedure and the subsequent trajectories used to obtain the IR spectrum

and dielectric constant. All simulation data are available at Github.[S13]



\end{document}